\documentclass[11pt]{article}
\usepackage[sc]{mathpazo} %Like Palatino with extensive math support
\usepackage{fullpage}
\usepackage[square,numbers]{natbib}
\usepackage[utf8]{inputenc}
\usepackage{lineno}

\usepackage{lineno}
\usepackage{amssymb}
\usepackage{amsfonts}
\usepackage{graphicx}
\usepackage{amsmath}
\usepackage{float,color,ulem}
\usepackage{epsf,epsfig,subfigure,diagbox}
\usepackage[toc,page,title,titletoc,header]{appendix}
\usepackage{indentfirst}

\usepackage[colorlinks=true]{hyperref}
\hypersetup{urlcolor=blue,linkcolor=black,citecolor=black,colorlinks=true}

\makeatletter\@addtoreset{equation}{section}
\newfloat{figure}{H}{lof}
\newfloat{table}{H}{lot}
\floatname{figure}{\figurename}
\floatname{table}{\tablename}
\newtheorem{theorem}{Theorem}[section]

\newtheorem{lemma}[theorem]{Lemma}

\newtheorem{remark}[theorem]{Remark}

\numberwithin{equation}{section}

\title{\textbf{Competition for light in forest population dynamics: from computer simulator to mathematical model}}
\author{Pierre Magal $^{1}$ \\
Zhengyang Zhang $^{1}$}

\date{}

\begin{document}

\maketitle

\noindent{}1. Univ. Bordeaux, IMB, UMR 5251, F-33076 Bordeaux, France; CNRS, IMB, UMR 5251, F-33400 Talence, France.

%Correspongding author P. Magal

\bigskip

%\textit{Manuscript elements}: Figure~1, figure~2, table~1, online
%appendices~A and B (including figure~A1 and figure~A2). Figure~2 is to
%print in color. 

%\bigskip


\bigskip

\textit{Manuscript type}: Article.

\linenumbers{}
\modulolinenumbers[3]

\newpage{}

\section*{Abstract}

In this article we build a mathematical model for forest growth and we compare this model with a computer forest simulator named SORTIE. The main ingredient taken into account in both models is the competition for light between trees. The parameters of the mathematical model are estimated by using SORTIE model, when the parameter values of SORTIE model correspond to the ones previously evaluated for the Great Mountain Forest in USA. We construct a size structured population dynamics model with one and two species and with spatial structure.

\newpage

\section*{Introduction}

In the natural ecosystem, forests play an important role. This has motivated a lot of people to propose computer simulators as well as mathematical models to describe the dynamical properties of forests. Many computer simulators (also sometimes called Individual Based Models (IBMs)) have been proposed and we refer to JABOWA (\citep{Botkin1972, Botkin1993}), FORET \citep{Shugart1977}, SORTIE \citep{Pacala1993}, FORMIND \citep{Koehler1998} and others. These models consist of stochastic processes describing individual behaviors, such as birth, death, movement, reproduction and so on. Moreover, these models also permit to describe the behavior of the entire plant community. The main advantage is that it provides simulated data which can be used to analyze such a complex system. Of course this is a rough description of the real plant community. However, they do supply powerful experimental tools and describe the forest dynamics reasonably well \citep{Levin1997}. We refer to \citep{Liu1995, Porte2002} for a general review about forest IBMs.

SORTIE is a forest simulator based on the forest data observed in and around Great Mountain Forest (GMF), a privately owned 2500ha forest located in northwestern Connecticut (41$^{\circ}$57'N, 73${^\circ}$15'W), USA in the year 1990-1992. In SORTIE, four submodels (resource, growth, mortality, and recruitment) are included to determine the behaviour of each individual. As is explained in \citep{Pacala1993}, SORTIE includes only the light limitation as the competition for resources, since little evidence of the effects of water or nitrogen on the growth has been found after extra experiments are performed. Tree growth is described by change of tree size, which is denoted here as the diameter at a certain height. Two concepts "$\mathrm{diam}_{10}$ (Diameter at 10cm Height)" and "DBH (Diameter at Breast Height)" are often used to describe the tree growth and represent the tree size in the analysis of forest dynamics (\citep{Pacala1993, Ribbens1994, Pacala1996}). Thereinto, the $\mathrm{diam}_{10}$ can be used almost throughout the whole life of an individual, from seedling to adult, while DBH can only be used for adults in most cases, as it is measured at a higher height. The definition of the breast height (of an adult human being) is different in different regions, for example, 1.4m in the US and 1.3m in Europe and Canada. But it makes little difference to the measuring result in many cases. We refer to \citep{Teck1991, Pacala1993, Pacala1994, Ribbens1994, Kobe1995, Pacala1996} for more details of SORTIE.

In this article we will extend the model proposed by Hal Smith in \citep{Smith1993, Smith1994} to describe the dynamic of a population that is structured in size with intra-specific competition for light. For a single species, we will compare such a mathematical model with SORTIE model for two types of tree (American beech (FAGR) and eastern hemlock (TSCA)). Moreover, based on the parameters estimated separately for each kind of tree, we will investigate the inter-specific competition for light by assuming that the growth rate is influenced by the competition for light. We will also extend our modelling effort by considering the case of two populations distributed in space and competing for light. 

Several mathematical models describing the forest growth were proposed in the literature. Zavala et al. \citep{Zavala2007} studied a stage-structured population model incorporating the light competition respectively in growth, mortality and recruitment, and gave the conditions for the existence of a steady state distribution. Angulo et al. \citep{Angulo2013} continued with a similar model, but considering the light competition only in recruitment, and they extended it to a two-species stand, and gave the positive stationary distribution for both single-species and two-species model, and the conditions for the coexistence. Cammarano \citep{Cammarano2011} studied a system of Lotka-Volterra type, incorporating also the light competition and discussed the equilibriums and the coexistence conditions. In this article, based on SORTIE simulated data, we will exclude the competition occurring in the mortality and recruitment. In other words, we will see that the best fit for SORTIE model is obtained by using a model where the competition for light influences only the growth rate of trees. We also refer to \citep{Roos2003, Strigul2008, Kohyama2012, Obiang2014} and the references therein for more kinds of models and researches.

The article is organised as follows. In section 2 we will give a mathematical model for single species, and we will conduct numerical simulations to compare with SORTIE. In section 3 a mathematical model for two species is obtained likewise, and we also conduct the comparison with SORTIE. Then in section 4 we extend it to a 2-dimension spatial model, and conduct numerical simulations to see the spread of trees in space.

\section*{Single species model}

\subsection*{Mathematical modelling}

In this section we consider the following model describing the growth of trees of single species
\begin{equation}\label{EQ2.1}
\left\{
\begin{array}{l}
\partial _{t}u(t,s)+\underset{\text{growth of adults}}{\underbrace{f(A(t))\partial _{s}u(t,s)}}=-\underset{\text{mortality}}{\underbrace{\mu (s)}}u(t,s),\text{ for } t>0,s>s_{-}, \vspace{0.1cm} \\
f(A(t))u(t,s_{-})=\underset{\text{flux of newborns}}{\underbrace{\beta b(A(t))}},\text{ for }t>0,\vspace{0.1cm} \\
u(0,.)=u_{0}(.)\in L^{1}(0,+\infty ).
\end{array}%
\right.  
\end{equation}
Here $u(t,s)$ denotes the population density of trees with size $s$ at time $t$, so $\displaystyle\int _{s_{1}}^{s_{2}}u(t,s)ds$ is the number of trees with size $s\in \lbrack s_{1},s_{2}]$ at time $t$, and $A(t)$ is the number of adult population at time $t$. The size $s$ is described by a function of $\mathrm{diam}_{10}$, which we will see in the appendix.  

The function $\mu (s)>0$ is the natural mortality. The minimal size of a juvenile is denoted as $s_{-}$. The parameter $\beta $ is the birth rate in absence of birth limitation, and the term $\beta b(A(t))$ describes the flux of newborns into the population, where $b(x)=xe^{-\xi x}$ is the Ricker's type birth limitation (\citep{Ricker1954, Ricker1975}). The growth function $f(x)$ takes the form
\begin{equation}\label{EQ2.2}
f\left( x\right) =\frac{\alpha }{1+\delta x},\ \alpha ,\delta >0,
\end{equation}%
which is decreasing, thus taking care of the fact that the more large trees there are, the slower the growth rate of small trees is. So this shows the type of competition for light between adults and juveniles. The function $u_{0}(.)$ represents the initial distribution of the species. Normally we want the number of the total population to be finite at each time, hence we have
\begin{equation*}
\int\nolimits_{0}^{+\infty }u(t,s)ds<+\infty ,\ \forall t\geqslant 0.
\end{equation*}%
So the natural state space for this model is $L^{1}$.

We will derive the following equations for adults and juveniles under some assumptions:

\begin{equation}\label{EQ2.3}
\left\{
\begin{array}{l}
\dfrac{dA(t)}{dt}=f(A(t))j(t,s^{\ast })-\mu _{A}A(t), \text{ for } t>0, \vspace{0.05cm} \\
\partial _{t}j(t,s)+f(A(t))\partial _{s}j(t,s)=-\mu _{J}j(t,s), \text{ for } s\in \lbrack s_{-},s^{\ast }),t>0, \vspace{0.05cm} \\
f(A(t))j(t,s_{-})=\beta b(A(t)), \text{ for } t>0, \vspace{0.05cm} \\
A(0)=A_0\geqslant 0, \\
j(0,s)=j_0(s)\geqslant 0,\text{ for } s\in \lbrack s_{-},s^{\ast }),
\end{array}
\right.
\end{equation}%
where $j(t,s)$ represents the population density of juveniles with size $s\in \lbrack s_{-},s^{\ast })$ at time $t$, and the positive constant $s^{\ast }$ satisfying $s^{\ast }>s_{-}$ denotes the maximal size of a juvenile (or the minimal size of an adult). Hence the total number of juveniles at time $t$ is
\begin{equation*}
J(t)=\int\nolimits_{s_{-}}^{s^{\ast
}}j(t,s)ds=\int\nolimits_{s_{-}}^{s^{\ast }}u(t,s)ds.
\end{equation*}
And we can assume as follows the adult population number
\begin{equation}\label{EQ2.4}
A(t)=\int\nolimits_{s^{\ast }}^{+\infty }u(t,s)ds.
\end{equation}
By integrating along the characteristic line of the second equation (of juvenile) in \eqref{EQ2.3}, the first equation (of adult) in (\ref{EQ2.3}) can be rewritten as the following state dependent Functional Differential Equation (FDE)
\begin{equation}\label{EQ2.5}
\left\{
\begin{array}{l}
\dfrac{dA(t)}{dt}=e^{-\mu _{J}\tau (t)}\dfrac{f(A(t))}{f(A(t-\tau (t)))}\beta b(A(t-\tau (t)))-\mu _{A}A(t),\\
\displaystyle\int\nolimits_{t-\tau (t)}^{t}f(A(\sigma ))d\sigma =s^{\ast }-s_{-}
\end{array}%
\right.  
\end{equation}
when $t>t^*$, where $t^*$ is defined as
\begin{equation*}
\int\nolimits_{0}^{t^{\ast }}f(A(\sigma ))d\sigma =s^{\ast }-s_{-}.
\end{equation*}%
Differentiation of the second equation with respect to $t$ gives the following system
\begin{equation}\label{EQ2.6}
\left\{
\begin{array}{l}
A^{\prime }(t)=e^{-\mu _{J}\tau (t)}\dfrac{f(A(t))}{f(A(t-\tau (t)))}\beta b(A(t-\tau (t)))-\mu _{A}A(t), \\
\tau ^{\prime }(t)=1-\dfrac{f(A(t))}{f(A(t-\tau (t)))}.
\end{array}%
\right.  
\end{equation}%
The initial conditions are
\begin{equation}\label{EQ2.7}
A(t)=A_{0}(t)\geqslant 0,\forall t\in (-\infty ,0];\ \tau (0)=\tau _{0}\geqslant 0,
\end{equation}
where $A_{0}(t)$ is continuous and exponentially bounded, namely for some $\vartheta >0$
\begin{equation*}
\sup_{t\leqslant 0}e^{\vartheta t}A_{0}(t)<+\infty .
\end{equation*}%
From the second equation of (\ref{EQ2.5}), as $f$ is decreasing, the delay $\tau(t)$ can become large enough, namely we may have infinite delay. For all the derivation here, see Appendix \ref{AppA}.

\subsection*{Numerical simulations of two special cases}

\begin{figure}[H]
\begin{minipage}{0.5\textwidth}
\begin{center}
\includegraphics[width=0.9\textwidth]{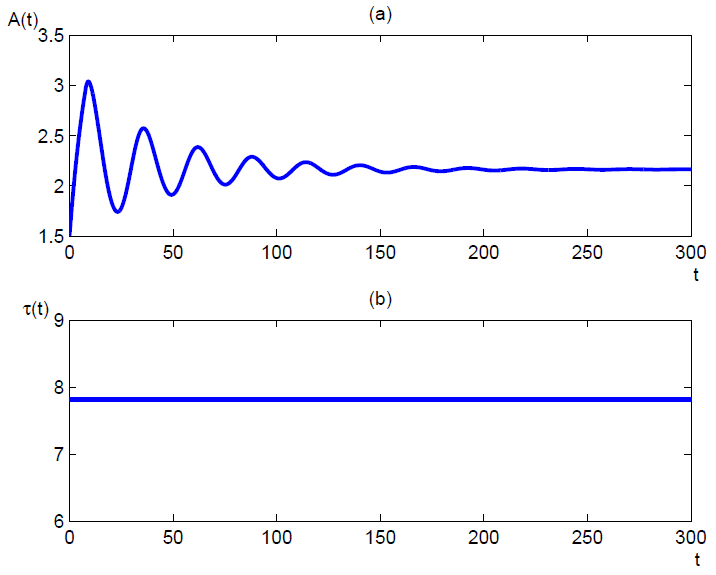}
\end{center}
\end{minipage}
\begin{minipage}{0.5\textwidth}
\begin{center}
\includegraphics[width=0.9 \textwidth]{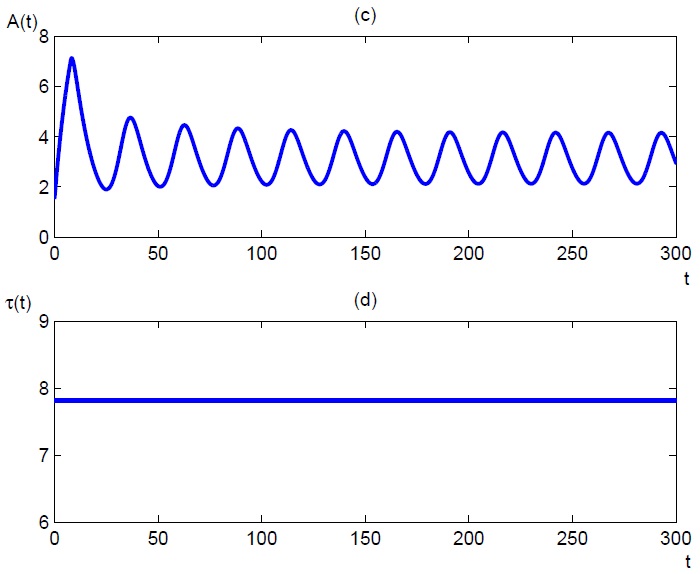}
\end{center}
\end{minipage}
\caption{\textit{We plot the adult population number $A(t)$ in figure (a) and (c), and the corresponding delay $\tau(t)$ in figure (b) and (d). We fix the parameter values $\mu_J=0.2$, $\mu_A=0.1$, $\alpha=0.5$, $\delta=0$, $\xi=1$, $s^*=\ln50$, $s_-=0$, and the initial distribution $\phi(t)=1.5, \forall t\in[-100,0]$. In (a) and (b) we set $\beta=9$. The solution oscillates and then converges to the positive equilibrium. In (c) and (d) we set $\beta=25$. Changing $\beta$ from 9 to 25, we observe a Hopf bifurcation.}}
\label{FIG5}
\end{figure}

We conduct numerical simulations for two special cases of the system \eqref{EQ2.5}.\\
\noindent \textbf{Special case 1 (}$f(x)$\textbf{\ is constant): }Assume that $f(x)$ is a constant function (so the delay $\tau(t)$ is also constant by the second equation of \eqref{EQ2.5}) and $b(x)=xe^{-x}$. Then since the PDE model \eqref{EQ2.1} can be transformed (by making a simple change of variable in time) into an age structured model, it is known (see Magal and Ruan \citep{Magal2009}) that the system has a Hopf bifurcation around the positive equilibrium when $\beta $ increases (see in Figure \ref{FIG5}).% shows that when $\beta $ passes from $9$ to $25$, we observe a Hopf bifurcation around the positive equilibrium.

\begin{figure}[H]
\begin{minipage}{0.5\textwidth}
\begin{center}
\includegraphics[width=0.9\textwidth]{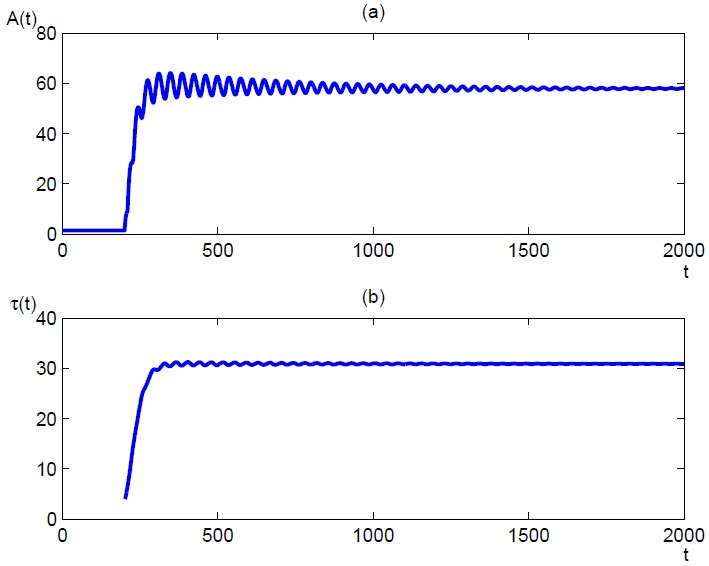}
\end{center}
\end{minipage}
\begin{minipage}{0.5\textwidth}
\begin{center}
\includegraphics[width=0.9\textwidth]{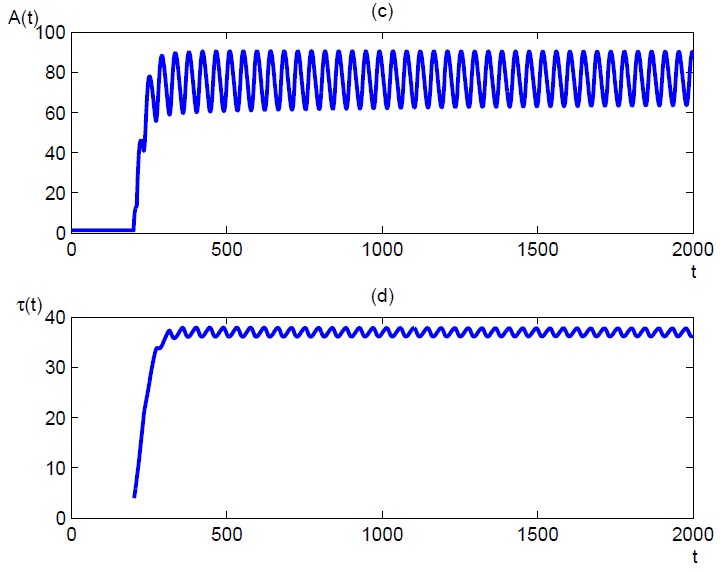}
\end{center}
\end{minipage}
\caption{\textit{We plot the adult population number $A(t)$ in figure (a) and (c), and the corresponding delay $\tau(t)$ in figure (b) and (d). We fix the parameter values $\mu_J=\mu_A=0.1$, $\delta=0.1$, $\xi=0$ (remember this means that $b(x)=x$), and the initial delay $\tau_0=4$. The initial distribution is $\phi(t)=1.5, \forall t\in[0,200]$. In (a) and (b) we set $\beta=2.2$, then we have the damped oscillating solution which converges to the positive equilibrium; In (c) and (d) we set $\beta=4$. Changing $\beta$ from 2.2 to 4, we observe a Hopf bifurcation.}} %The delay is shown in figure (b) and (d), which is also periodic, but with a relatively small amplitude.}}
\label{FIG6}
\end{figure}

\noindent \textbf{Special case 2 (}$b(x)=x$\textbf{): }Assume $b(x)=x$, namely the Ricker's type birth function doesn't appear in system \eqref{EQ2.5}. It is known that when $\tau$ is constant in the first equation (which becomes linear) of system \eqref{EQ2.5}, this system is either exponentially increasing or exponentially decreasing when the time goes to infinity. However, it has been proved by Smith \citep{Smith1993} that Hopf bifurcation can occur when we take state-dependent delay. This is illustrated in Figure \ref{FIG6}.

\subsection*{Comparison with SORTIE}

We run the simulator SORTIE with the parameter values given in \citep{Teck1991, Pacala1993, Ribbens1994, Kobe1995} and get the simulation for the density of adult trees %in Figure \ref{FIG2} 
(adults are defined here as trees having a DBH $\geqslant$ 10cm). And as we can see from this simulation, American beech(FAGR) and eastern hemlock(TSCA) %(in Figure \ref{FIG3}) 
become the dominant species after a period time. So in this article we will focus on these two species in two cases: one single species and two-species.

The basic idea of the numerical simulation of (\ref{EQ2.5}) and comparison is as follows. Before starting, we need to get the forest data from SORTIE. Since every run of SORTIE is initiated with a random seed, we conduct 50 runs and take the average values as our actual data. Moreover, the data that SORTIE gives are actually the density of the adult population per hectare. As the area of the sample square is $90000\mathrm{m}^2$ (a square of 300m$\times$300m) = 9 hectares, we multiply the data by 9 to obtain the total adult population number. We plot them and the average in MATLAB.% to see the variation in Figure \ref{FIG7}.

%\begin{figure}[H]
%\begin{minipage}{0.5\textwidth}
%\begin{center}
%{\bf (a)}
%\includegraphics[width=0.7\textwidth]{variabfagr.jpeg}
%\end{center}
%\end{minipage}
%\begin{minipage}{0.5\textwidth}
%\begin{center}
%{\bf (b)}
%\includegraphics[width=0.7\textwidth]{variabtsca.jpeg}
%\end{center}
%\end{minipage}
%\begin{minipage}{0.5\textwidth}
%\begin{center}
%{\bf (c)}
%\includegraphics[width=0.7\textwidth]{variabfagr2.jpeg}
%\end{center}
%\end{minipage}
%\begin{minipage}{0.5\textwidth}
%\begin{center}
%{\bf (d)}
%\includegraphics[width=0.7\textwidth]{variabtsca2.jpeg}
%\end{center}
%\end{minipage}
%\caption{\textit{Population numbers of 50 random runs from SORTIE for the adults of two species American beech and eastern hemlock(green curve) and the average number(red curve) respectively, in a period of 400 timesteps. (a). One single species American beech; (b). One single species eastern hemlock; (c). Two-species case: American beech; (d). Two-species case: eastern hemlock.}}
%\label{FIG7}
%\end{figure}

Now we will compare our model (\ref{EQ2.5}) with the mean value over these 50 runs of SORTIE, %As the timestep in SORTIE is 5 years, we conduct a linear interpolation, changing the timestep to 1 year, in order to make the simulation more accurate.
and find the best fit. First we need to decide the initial time (for example, $t=100$ as the initial time), and we will use the data from SORTIE over the time interval $[0,100]$ as the initial condition. %In fact, in order to make our simulation closer to the data, we use a part of the data in the beginning given from SORTIE (for example, the data in the time interval [0,100]) as the initial value for the delay differential equation. We believe that using this initial distribution will lead to a better fit. 
Next we discretize the parameters $\mu _{J},\mu _{A},\beta ,\xi ,\delta ,\tau _{0}$, and for each set of parameters, we calculate the solution of (\ref{EQ2.5}) by using the common approximation of the derivative (the numerical scheme will be conducted via the equivalent system (\ref{EQ2.6})), and we compare the numerical solutions with the data from SORTIE by using the least square method, to find the set of parameter values with which the numerical result of the model (\ref{EQ2.5}) and the data have the least difference. Then we use the following formula (see Appendix \ref{AppA})%(from (\ref{EQ2.14})):
\begin{equation*}
\int_{-\tau_{0}}^0\frac{\alpha}{1+\delta A(\sigma)}d\sigma=s^*-s_-,
\end{equation*}
to compute $\alpha$, where we use the Simpson's rule to calculate the integral. Now we can keep this set of parameter values, and we have the best fit to SORTIE.

\begin{figure}[H]
\begin{center}
\includegraphics[width=0.6\textwidth]{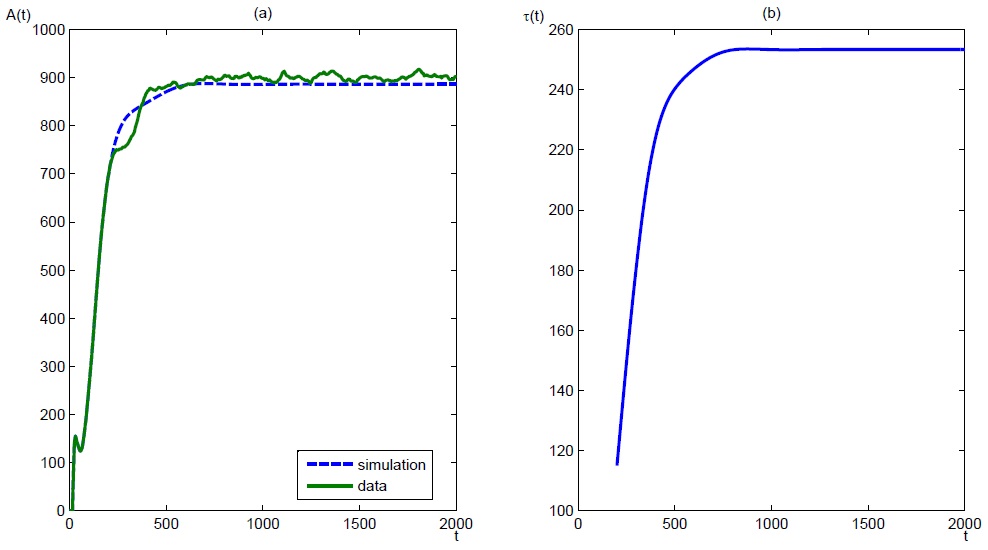}
\end{center}
\caption{\textit{In this figure we show the comparison between SORTIE data and numerical simulation for American beech. The adult population number $A(t)$ is plotted in (a) and the delay $\tau(t)$ is shown in (b).}} %Notice that the solution increases and converges to the stable positive equilibrium.
\label{FIG8}
\end{figure}

For the first dominant species American beech, we choose the SORTIE data in the time interval [0,200] as the initial distribution. We have the best fit in Table \ref{TABLE1} and Figure \ref{FIG8}.

%The parameter value $\xi_1 =0$ means that $b(x)=x$, namely there is no Ricker's type birth limitation here.

\begin{figure}[H]
\begin{center}
\includegraphics[width=0.6\textwidth]{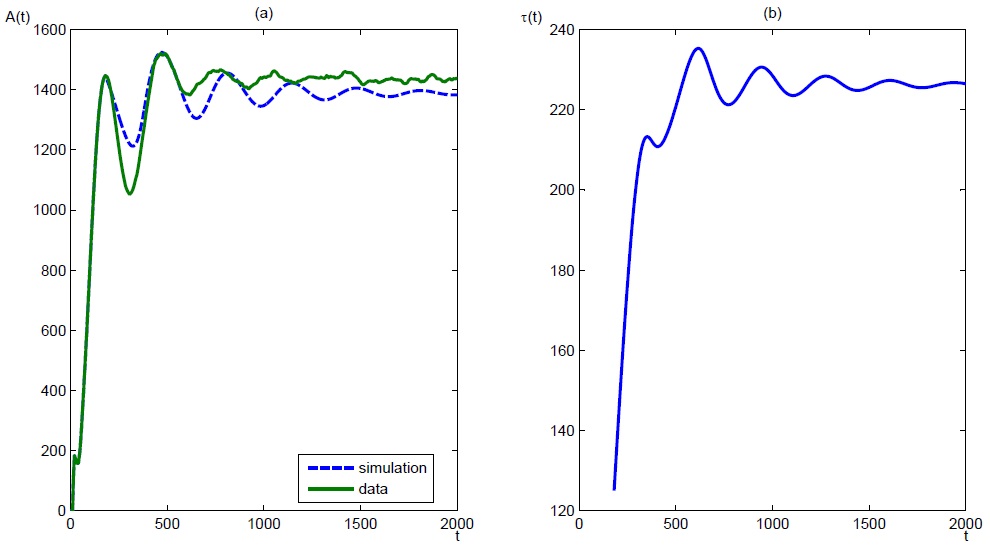}
\end{center}
\caption{\textit{In this figure we show the comparison between SORTIE data and numerical simulation for eastern hemlock. The adult population number $A(t)$ is plotted in (a) and the delay $\tau(t)$ is shown in (b). Notice that this species will have an oscillation before it converges to the stable positive equilibrium.}}
\label{FIG9}
\end{figure}

Similarly, for the second dominant species eastern hemlock we choose the SORTIE data in the time interval [0,180] as the initial distribution. We get the best fit in Table \ref{TABLE2} and Figure \ref{FIG9}.

Notice that for both species in the single species case, we have $\xi_{1}=\xi_{2}=0$ as the best fit, which means that there is no Ricker's type birth limitation here.

\section*{Two-species model}

\subsection*{Mathematical modelling}

System (\ref{EQ2.1}) can be extended to the case of two species. Taking the previous best fit $\xi _{1}=\xi _{2}=0$ into account, we obtain the following system

\begin{equation}\label{EQ3.1}
\left\{
\begin{array}{l}
\partial _{t}u_{1}(t,s)+f_{1}(Z_{1}(t))\partial _{s}u_{1}(t,s)=-\mu _{1}(s)u_{1}(t,s), \text{ for } t>0,s>s_{-}, \\
\partial _{t}u_{2}(t,s)+f_{2}(Z_{2}(t))\partial _{s}u_{2}(t,s)=-\mu _{2}(s)u_{2}(t,s), \text{ for } t>0,s>s_{-}, \\
f_{1}(Z_{1}(t))u_{1}(t,s_{-})=\beta_1A_1(t), \text{ for } t>0, \\
f_{2}(Z_{2}(t))u_{2}(t,s_{-})=\beta_2A_2(t), \text{ for } t>0, \\
u_{1}(0,\cdot )=u_{10}(\cdot )\in L^{1}(0,+\infty ), \\
u_{2}(0,\cdot )=u_{20}(\cdot )\in L^{1}(0,+\infty ),%
\end{array}%
\right.
\end{equation}
where 
\begin{equation*}
Z_{i}(t)=\zeta _{i1}A_{1}(t)+\zeta _{i2}A_{2}(t),\ f_{i}(x)=\frac{\alpha _{i}}{1+\delta _{i} x},\ \mu _{i}(s)=\left\{\begin{array}{ll}
\mu_{A_i}>0, & \text{if}\ s\geqslant s^*,\\
\mu_{J_i}>0, & \text{if}\ s\in [s_-,s^*),
\end{array}\right.
\end{equation*}
and $\zeta_{ij}\geqslant0$ are nonnegative constants, $\alpha _{i},\delta _{i}>0$, $i,j=1,2$. Notice that we use the same minimal juvenile size $s_-$ and minimal adult size $s^*$ for both species(\citep{Pacala1993}). After a similar derivation, we have the following state-dependent delay differential equations
\begin{equation}\label{EQ3.2}
\left\{
\begin{array}{l}
A_{i}^{\prime }(t)= e^{-\mu _{J_{i}}\tau _{i}(t)}\dfrac{f_{i}(Z_{i}(t))}{f_{i}(Z_{i}(t-\tau _{i}(t)))}\beta _{i}A_{i}(t-\tau _{i}(t))-\mu _{A_{i}}A_{i}(t), \\
\displaystyle\int_{t-\tau_i(t)}^t f_i(Z_{i}(t))d\sigma=s^*-s_-,
\end{array}
\right.
\end{equation}
$i=1,2$. We give the following expression for the sake of numerical simulation:
\begin{equation}\label{EQ3.3}
\left\{
\begin{array}{l}
A_{i}^{\prime }(t)= e^{-\mu _{J_{i}}\tau _{i}(t)}\dfrac{f_{i}(Z_{i}(t))}{f_{i}(Z_{i}(t-\tau _{i}(t)))}\beta _{i}A_{i}(t-\tau _{i}(t))-\mu _{A_{i}}A_{i}(t), \vspace{0.1cm} \\
\tau _{i}^{\prime }(t)=1-\dfrac{f_{i}(Z_{i}(t))}{f_{i}(Z_{i}(t-\tau _{i}(t)))}.
\end{array}%
\right.
\end{equation}

\subsection*{Comparison with SORTIE}

\begin{figure}[H]
\begin{center}
\includegraphics[width=0.6\textwidth]{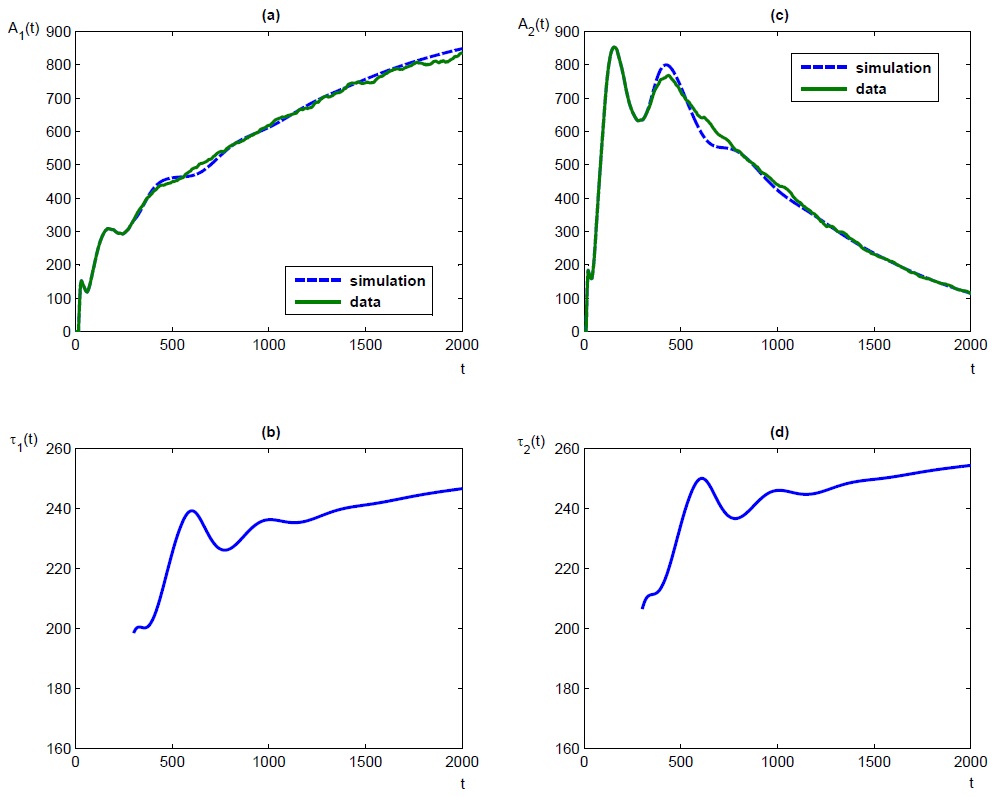}
\end{center}
\caption{\textit{In this figure we plot the comparison between SORTIE data and numerical simulation for two-species model (\ref{EQ3.2}). The figures (a) and (c) show the adult population number for species 1 American beech and species 2 eastern hemlock respectively. The figures (b) and (d) show the corresponding time delay.}}
\label{FIG10}
\end{figure}

We use the same method of comparison as before and we use the parameters in Table \ref{TABLE1} and Table \ref{TABLE2} to simulate the two-species model. We discretize the new parameters $\zeta_{ij}$ appeared in the competition term, and also by the least square method, we get the best fit for them: $\zeta_{11}=1$, $\zeta_{12}=0.6$, $\zeta_{21}=1.6$, $\zeta_{22}=1$, and the delay $\tau_{01}$ is increased to 201, $\tau_{02}$ increased to 208. We list these values in Table \ref{TABLE3}. The comparison figure is in Figure \ref{FIG10}.

By analyzing the existence positive (coexisting) equilibrium (see Appendix \ref{AppB}), we can also obtain the coexistence of both American beech and eastern hemlock.  
\begin{figure}[H]
\begin{center}
\includegraphics[width=0.6\textwidth]{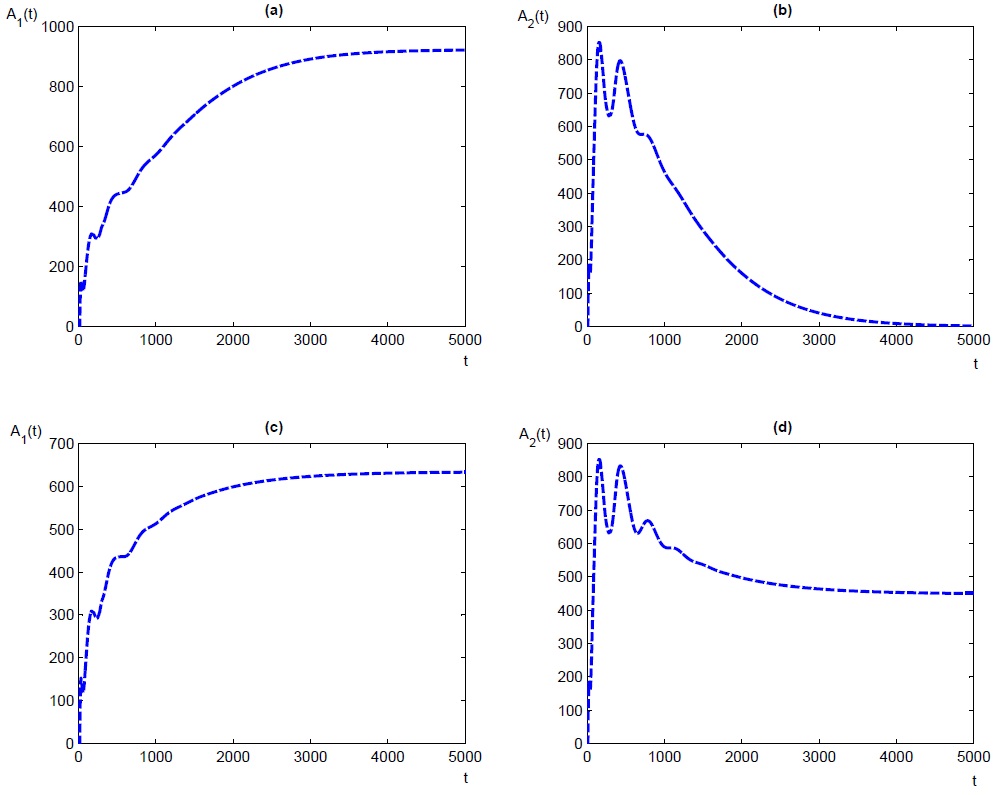}
\end{center}
\caption{\textit{In this figure we demonstrate that we can pass from competitive exclusion (a)(b)(which corresponds to Figure \ref{FIG10}) to coexistence (c)(d) by changing one parameter $\zeta_{21}$. The other parameters are the same as in Figure \ref{FIG10}. When $\zeta_{21}=1.6$, American beech (a) reaches to a positive steady state while eastern hemlock (b) disappears gradually. After we decrease the value of $\zeta_{21}$ to $1$, both American beech (c) and eastern hemlock (d) go to a positive steady state, which means coexistence.}}
\label{FIG11}
\end{figure}

\section*{Two-species spatial model}

Now we take the spatial position of the individuals into account to see the spread of the adult population. Inspired by Ducrot \citep{Ducrot2011}, we can describe the spreading of the seeds around the adult trees by using the following two species model 

\begin{equation}\label{EQ4.1}
\left\{\begin{array}{l}
\partial_t u_i(t,s,x,y)+f_i(Z_i(t,x,y))\partial_s u_i(t,s,x,y)=-\mu_i(s)u_i(t,s,x,y), \\
\qquad\qquad\qquad\qquad\qquad\qquad\qquad\text{ for } t>0,s>s_{-},x\in [0,x_{\max}],y\in [0,y_{\max}], \\
f_i(Z_i(t,x,y))u_i(t,s_-,x,y)=(I-\varepsilon_i\Delta)^{-1}(\beta_iA_i(t,.,.))(x,y), \\
\qquad\qquad\qquad\qquad\qquad\qquad\qquad\text{ for }t>0,x\in [0,x_{\max}],y\in [0,y_{\max}], \\
u_i(t,x,0)=u_i(t,x,y_{\max}),\text{ for } x\in[0,x_{\max}],\\
u_i(t,0,y)=u_i(t,x_{\max},y),\text{ for } y\in[0,y_{\max}],\\
u_i(0,s,x,y)=u_{i0}(s,x,y)\in L^1((0,+\infty)\times[0,x_{\max}]\times[0,y_{\max}]),
\end{array}\right.
\end{equation}
where
\begin{equation*}
Z_i(t,x,y)=\zeta_{i1}A_1(t,x,y)+\zeta_{i2}A_2(t,x,y),f_i(x)=\frac{\alpha_i}{1+\delta_i x},
\end{equation*}
$\zeta _{ij}\geqslant 0$, $\alpha _{i},\delta _{i}>0$, $i=1,2$, and $\Delta$ is the Laplacian operator with periodic boundary condition. Similarly, we assume the adult population number
\begin{equation*}
A_i(t,x,y)=\int_{s^*}^{+\infty}u_i(t,s,x,y)ds,i=1,2,
\end{equation*}
and by following a similar procedure as before, we get the state-dependent delay differential equation for the adult
\begin{equation}\label{EQ4.2}
\left\{\begin{array}{l}
\displaystyle\begin{aligned}\frac{\partial A_i(t,x,y)}{\partial t}= & e^{-\mu_{J_{i}}\tau_i(t,x,y)}\frac{f_i(Z_i(t,x,y))}{f_i(Z_i(t-\tau_i(t,x,y),x,y))}(I-\varepsilon_i\Delta)^{-1}(\beta_iA_i(t-\\
& \tau_i(t,x,y),.,.))(x,y)-\mu_{A_{i}}A_i(t,x,y),\text{ for } t>t^*,\end{aligned}\\
\displaystyle\int_{t-\tau_i(t,x,y)}^tf_i(Z_i(\sigma,x,y))d\sigma=s^*-s_-,\text{ for } t>t^*.
\end{array}\right.
\end{equation}
We conduct numerical simulations for system (\ref{EQ4.2}), using the parameters in Table \ref{TABLE1}-\ref{TABLE3}, and setting the diffusion coefficient $\varepsilon_1=0.01$, $\varepsilon_2=0.005$, in order to observe the growth and spread of adult population of the two species. The simulation is conducted in a $300*300$ square of the $x-y$ plane, as in the reference \citep{Pacala1993}. We set the random initial distribution for the two species by taking a random number at 25 random positions.% and we have one as follows: $\forall t\in[-200,0]$, 
%\begin{table}[H]
%\begin{center}
%\begin{tabular}{|l|ccccc|} 
% \hline 
% \diagbox{x}{$A_1$}{y} & 260 & 240 & 40 & 300 & 200 \\ 
% \hline 
% 280 & 3.5175 & 12.1478 & 5.3812 & 11.5242 & 12.9524 \\ 
% 105 & 14.4352 & 3.8349 & 15.3100 & 13.6673 & 13.5803 \\
% 15 & 9.4697 & 14.7685 & 3.7732 & 10.9319 & 12.7157 \\
% 250 & 3.0544 & 4.8570 & 5.7500 & 8.5146 & 18.9035 \\
% 180 & 6.8225 & 18.3485 & 1.8223 & 12.8889 & 4.1787 \\
% \hline 
% \end{tabular} 
%\end{center}
%\begin{center}
% \begin{tabular}{|l|ccccc|} 
% \hline 
% \diagbox{x}{$A_2$}{y} & 295 & 155 & 140 & 55 & 85 \\ 
% \hline 
% 175 & 10.6392 & 6.8809 & 6.2424 & 8.7337 & 1.7882 \\ 
% 70 & 3.5435 & 9.9292 & 12.6289 & 8.1111 & 14.0974 \\
% 35 & 1.7909 & 11.5543 & 12.4938 & 13.0491 & 9.6833 \\
% 110 & 9.1096 & 5.2533 & 3.8466 & 3.9717 & 7.1919 \\
% 200 & 6.7521 & 9.9301 & 9.2019 & 4.7711 & 9.5898 \\
% \hline 
% \end{tabular}
% \end{center}
% \caption{\textit{A random initial distribution for spatial model (\ref{EQ4.2}). This random initial distribution gives the population number $A_1(t)$ and $A_2(t)$ at a certain spot. For example, at the point with coordinate (280,260), we have $A_1(t,280,260)=3.5175$, $\forall t\in[-200,0]$. For the spots other than those in the tables, the values of $A_1$ and $A_2$ are zero.}}
% \end{table}

\begin{figure}[H]
\begin{minipage}{0.5\textwidth}
\begin{center}
\includegraphics[width=0.7\textwidth]{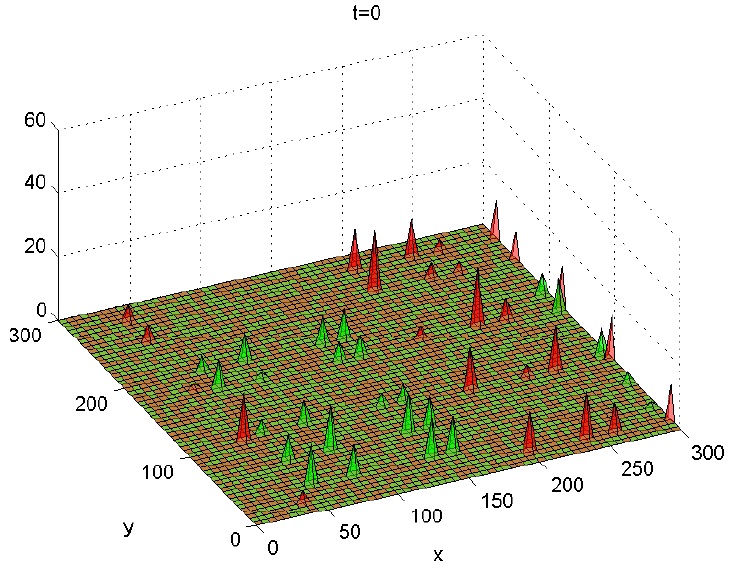}
\end{center}
\end{minipage}
\begin{minipage}{0.5\textwidth}
\begin{center}
\includegraphics[width=0.7\textwidth]{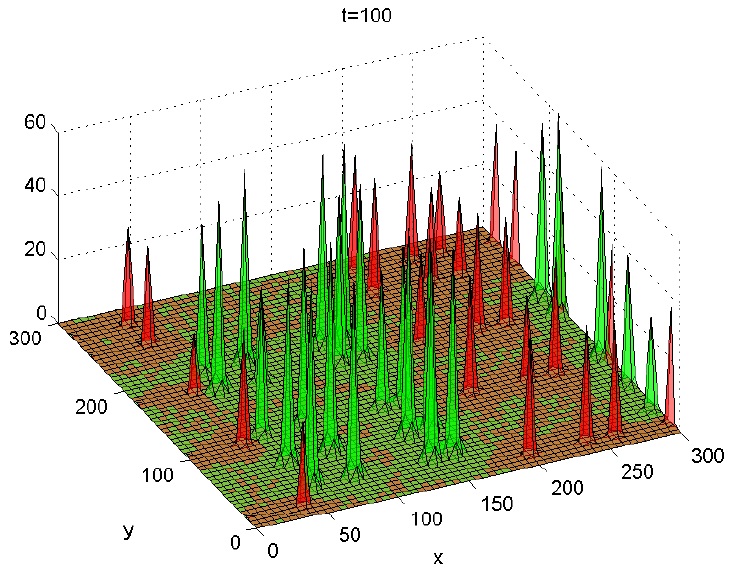}
\end{center}
\end{minipage}
\begin{minipage}{0.5\textwidth}
\begin{center}
\includegraphics[width=0.7\textwidth]{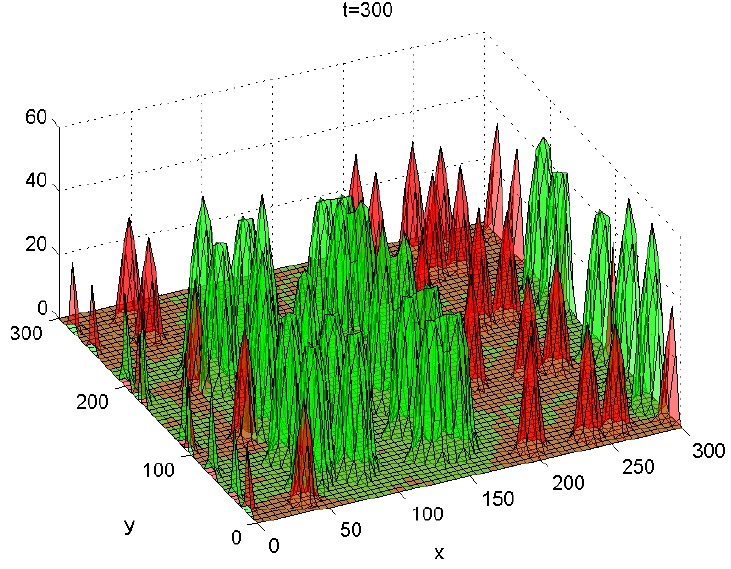}
\end{center}
\end{minipage}
%\begin{minipage}{0.5\textwidth}
%\begin{center}
%\includegraphics[width=0.7\textwidth]{500.jpeg}
%\end{center}
%\end{minipage}
%\begin{minipage}{0.5\textwidth}
%\begin{center}
%\includegraphics[width=0.7\textwidth]{700.jpeg}
%\end{center}
%\end{minipage}
\begin{minipage}{0.5\textwidth}
\begin{center}
\includegraphics[width=0.7\textwidth]{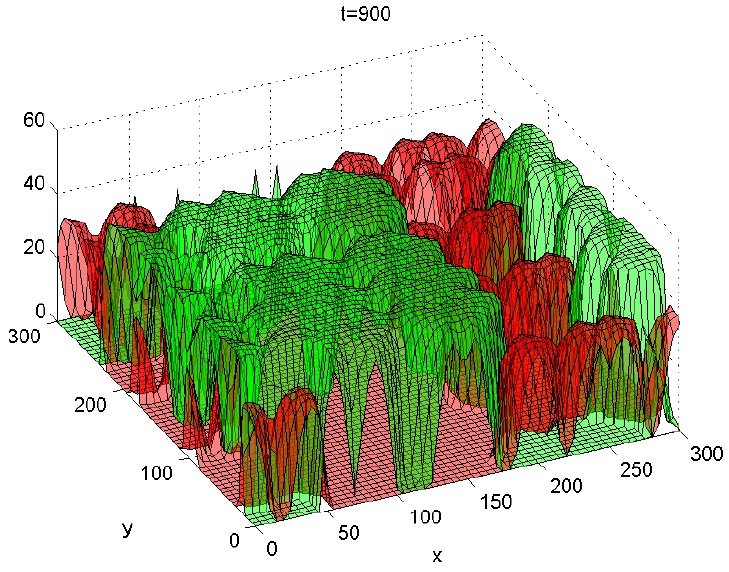}
\end{center}
\end{minipage}
\caption{\textit{In this figure we show the numerical simulations for the spatial model (\ref{EQ4.2}), describing the spread of adult population of the two species. The red part represents species 1 American beech, and the green part represents species 2 eastern hemlock.}}
\label{FIG12}
\end{figure}

Next we plot the solutions of system (\ref{EQ4.2}) at several specified time in Figure \ref{FIG12}. The $x$- and $y$-axis describe the spatial coordinates, and the $z$-axis is the adult population number. In this figure we will observe the growth of the two species and the spread in space, and we can also see vividly that the model generates obvious species isolates after some time.

\begin{figure}[H]
\begin{minipage}{0.5\textwidth}
\begin{center}
\includegraphics[width=0.7\textwidth]{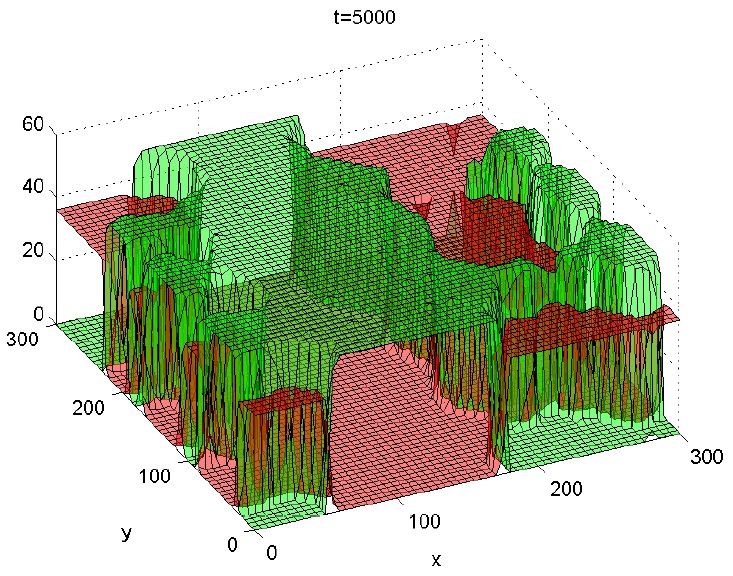}
\end{center}
\end{minipage}
\begin{minipage}{0.5\textwidth}
\begin{center}
\includegraphics[width=0.7\textwidth]{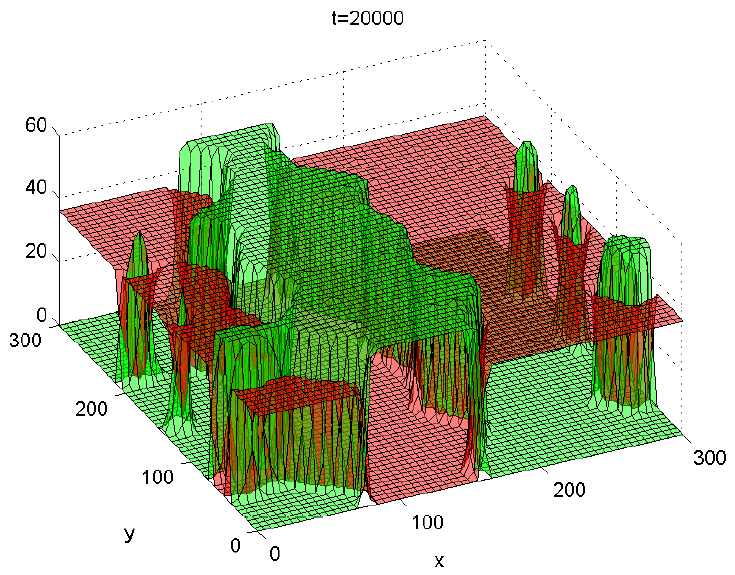}
\end{center}
\end{minipage}
\begin{minipage}{0.5\textwidth}
\begin{center}
\includegraphics[width=0.7\textwidth]{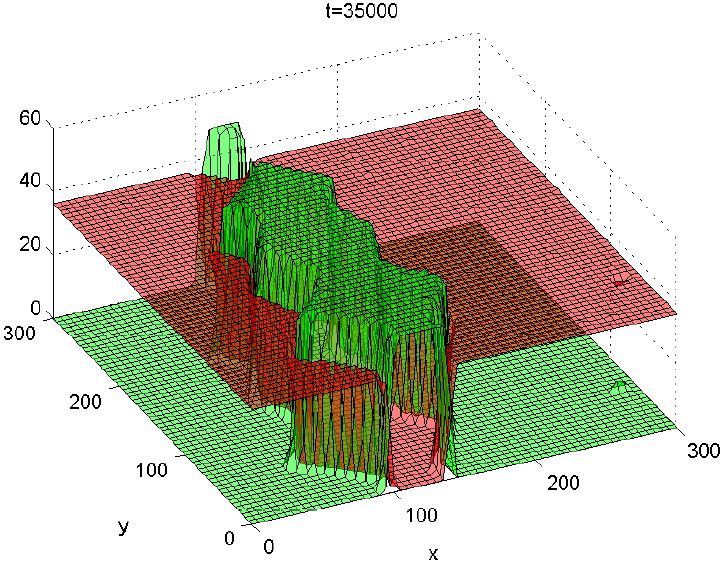}
\end{center}
\end{minipage}
\begin{minipage}{0.5\textwidth}
\begin{center}
\includegraphics[width=0.7\textwidth]{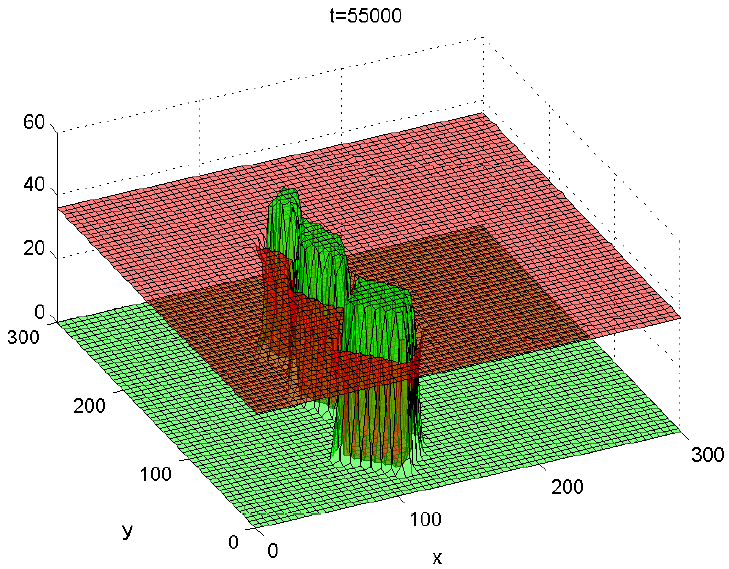}
\end{center}
\end{minipage}
\caption{\textit{This figure shows the change of the distribution of adult population of two species in the long run. Notice that eastern hemlock(green) is disappearing and American beech(red) becomes dominant.}}
\label{FIG13}
\end{figure}

We also conduct the simulation for longer time, and we get the following results in Figure \ref{FIG13}.

Summarizing all the figures above, we may conclude that eastern hemlock(green) grows and spreads faster than American beech(red) at first, but after long enough time, American beech begins to show its competency and gradually becomes the dominant species. This result also coincides with our previous result without considering the space in Figure \ref{FIG10}.

\begin{figure}[H]
\begin{center}
\includegraphics[width=0.4\textwidth]{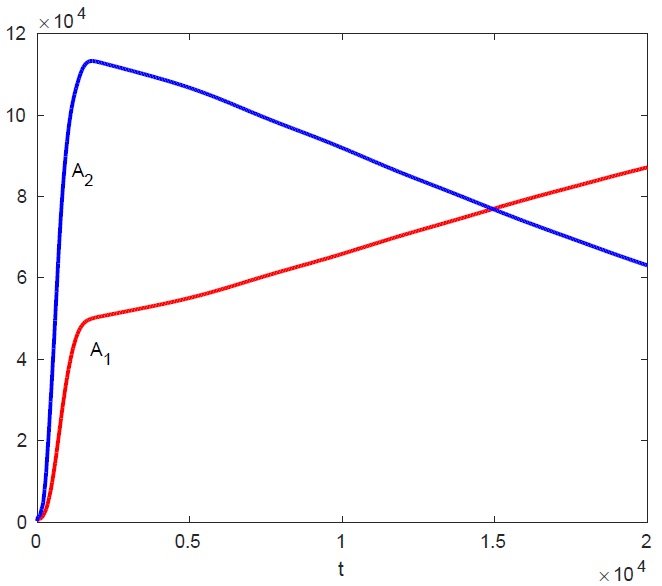}
\end{center}
\caption{\textit{In this figure we show the total population in the $300m \times 300 m$ square for each species in 20000 years. $A_1$ represents species 1 American beech, and $A_2$ represents species 2 eastern hemlock.}}
\label{FIG14}
\end{figure}

Moreover, we plot the total population in the sample square for each species with respect to time in Figure \ref{FIG14}. And we can see that the total adult population for eastern hemlock increases faster than American beech at first, and then it decreases.

\begin{figure}[H]
\begin{minipage}{0.5\textwidth}
\begin{center}
\includegraphics[width=0.7\textwidth]{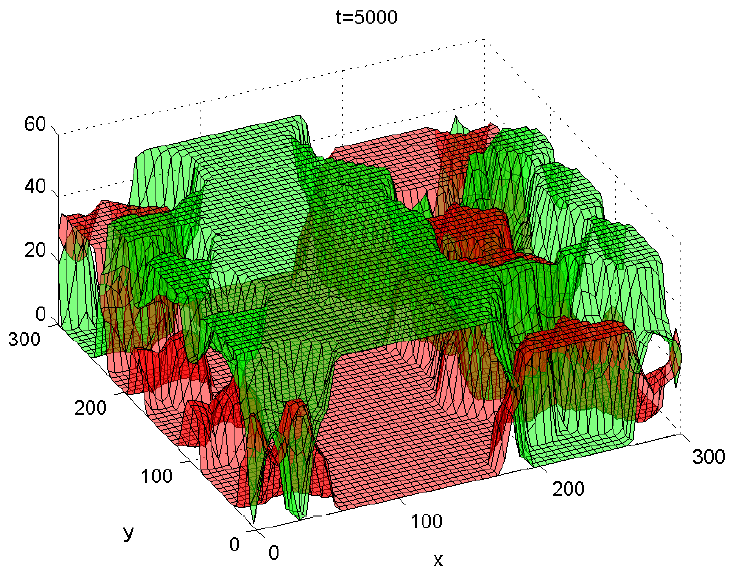}
\end{center}
\end{minipage}
\begin{minipage}{0.5\textwidth}
\begin{center}
\includegraphics[width=0.7\textwidth]{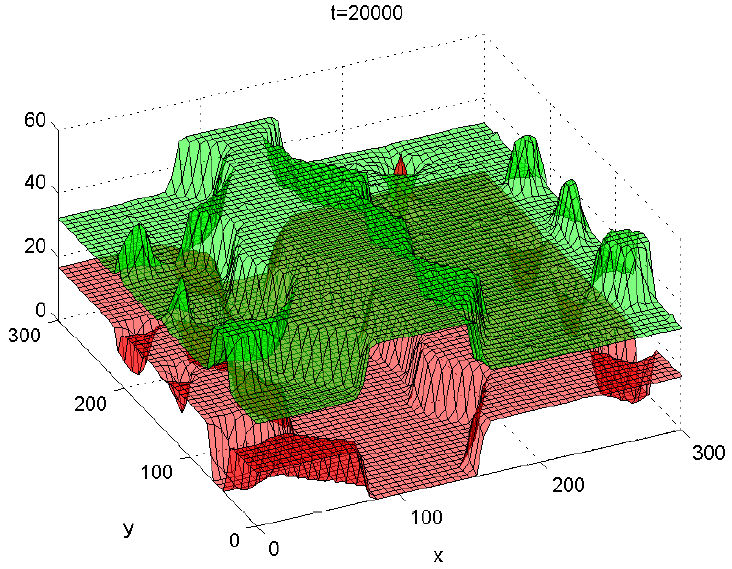}
\end{center}
\end{minipage}
\begin{minipage}{0.5\textwidth}
\begin{center}
\includegraphics[width=0.7\textwidth]{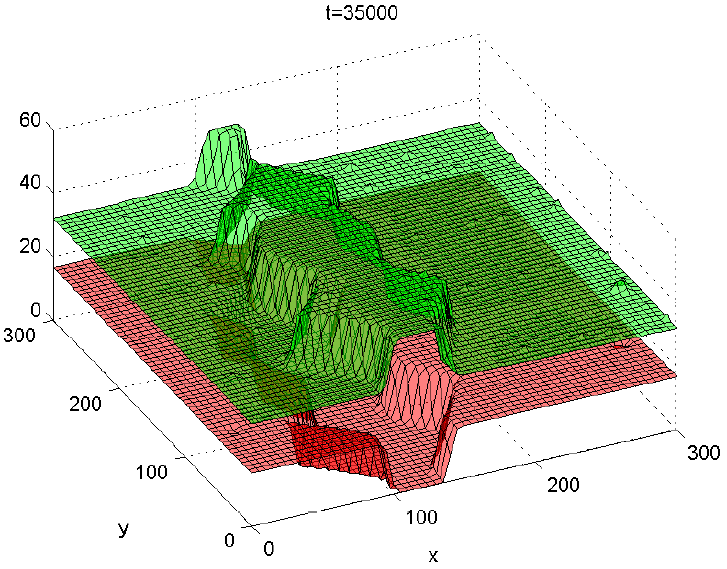}
\end{center}
\end{minipage}
\begin{minipage}{0.5\textwidth}
\begin{center}
\includegraphics[width=0.7\textwidth]{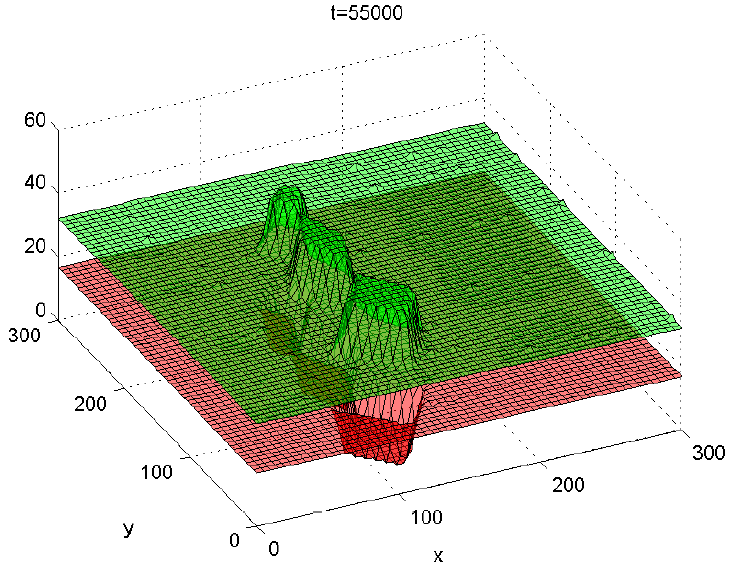}
\end{center}
\end{minipage}
\caption{\textit{The longterm simulation for the spatial model with a change of the parameter $\zeta_{21}$ from $1.6$ to $1$. All the other parameters are the same as in Figure \ref{FIG13}. We observe the coexistence of both species.}}
\label{FIG15}
\end{figure}

As in Figure \ref{FIG11}, we can also observe the coexistence of both species in the spatial model. In Figure \ref{FIG15} we plot the long term distributions after the same change of parameters as in Figure \ref{FIG11}. 

\section*{Discussion}

Studies of forest dynamics have a long history. There have been a large amount of researches on either the descriptive model for forests reached by observed data, or the pure mathematical model with numerical computations, separated from data. Here we first construct a mathematical model and compare this model to the computer forest simulator SORTIE. To our best knowledge there is no similar work in the literature. 

We start by fitting the parameters of the model by considering the case of a single species. Then for two species, we only fit the parameters corresponding to the competition for light between the two species of trees. Specifically speaking, we use a classical size-structured model, from which we derive a state-dependent delay differential equation, and we use this differential equation to fit the forest data from SORTIE. This differential equation is mathematically more tractable than the submodels in SORTIE. 

In order to compare our mathematical model with SORTIE, we conduct numerical simulations and we get the best fit to the SORTIE forest data and the corresponding parameter values. One result we get is that the type of birth function of these two species is not of Ricker's type, as we have $\xi=0$ in both best fits. We then extend our mathematical model to a two-species case with interspecific competition, and similarly we conduct the numerical comparison with SORTIE forest data, where we also get a very good fit.

Based on this, we go further and propose a model incorporating the spatial position parameter, to describe the density of population, or further, the number of population at every specific spatial position. We can see vividly the spread and succession in our numerical simulation. By using a spatial model, given the initial distribution we will be able to predict specifically the population at certain spatial position and time, which is more practical in reality. 

In our simulation, we reach a result that the population number of eastern hemlock decreases to 0 after a long enough time, which conforms to the competitive exclusion principle. However, by analyzing the existence of the interior coexistent equilibrium, we are able to establish a range of parameters in which the exclusion principle is no longer true. The coexistence result has also been confirmed by numerical simulations (with and without space). We refer to \citep{Zavala2007, Cammarano2011, Kohyama2012, Angulo2013} for more result going into that direction. But there are few results about the coexistence for the solution of the structured model. Also it is well known that light is a key influence in many forest systems (\citep{Valladares2008}), and our model can be used to reproduce the complicated mechanisms included into SORTIE model. But in reality, there are so many influencing factors, such as carbon, nitrogen, water, etc.(\citep{Kolb1990, Loustau1995, Kramer2000, Cheaib2005, Trichet2008}), not only restricted to light. A lot of work is left for future investigation.   

\begin{appendices}

\section{Derivation of the state dependent FDE}\label{AppA}

\begin{figure}[H]
\begin{center}
\includegraphics[width=0.6\textwidth]{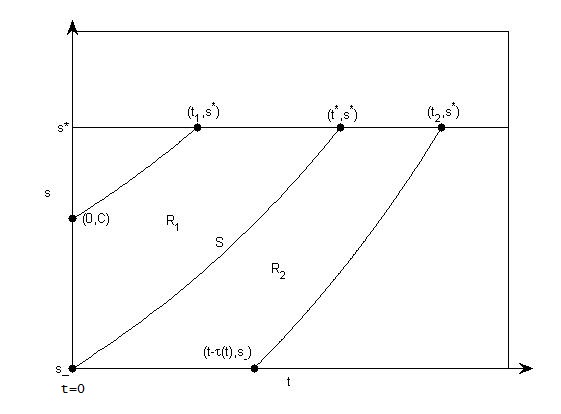}
\end{center}
\caption{\textit{In this figure we present the characteristic curves (\ref{EQ2.9}).}}
\label{FIG4}
\end{figure}

The single species model (\ref{EQ2.1}) we consider here is very similar with the one considered by H. Smith in \citep{Smith1993}. Nevertheless, our mortality rate $\mu(s)$ is dependent on the size $s$, so we will re-derive the state dependent FDE for completeness. Differentiating (\ref{EQ2.4}) with respect to $t$, we have
\begin{eqnarray}
\frac{dA(t)}{dt} &=&\int\nolimits_{s^{\ast }}^{+\infty }\partial
_{t}u(t,s)ds=-f(A(t))\int\nolimits_{s^{\ast }}^{+\infty }\partial
_{s}u(t,s)ds-\int\nolimits_{s^{\ast }}^{+\infty }\mu(s)u(t,s)ds  \notag \\
&=&f(A(t))u(t,s^{\ast })-\int\nolimits_{s^{\ast }}^{+\infty }\mu(s)u(t,s)ds. \label{EQ2.8}
\end{eqnarray}%
Next we deal with the term $u(t,s^{\ast })$. The characteristic curves for
the first equation in (\ref{EQ2.1}) are (shown in Figure \ref{FIG4})%
\begin{equation}\label{EQ2.9}
\frac{ds(t)}{dt }=f(A(t)).
\end{equation}
Then we will have the following representation of $s$
\begin{equation}\label{EQ2.10}
C+\int\nolimits_{0}^{t}f(A(\sigma ))d\sigma =s(t).  
\end{equation}%
Suppose $t^{\ast }$ is the time when juveniles present at time 0 become adults, namely
\begin{equation}\label{EQ2.11}
\int\nolimits_{0}^{t^{\ast }}f(A(\sigma ))d\sigma =s^{\ast }-s_{-}.
\end{equation}%
We can see that the curve
\begin{equation*}
S=\left\{ (t,s):\text{ }0\leqslant t\leqslant t^{\ast },\text{ }%
s=s_-+\int\nolimits_{0}^{t}f(A(\sigma ))d\sigma \right\}
\end{equation*}%
divides the strip $\left[ 0,+\infty \right) \times \left[ s_{-},s^{\ast }%
\right] $ into two parts $R_{1}$ and $R_{2}$.
Assuming that $s-s_{-} \leqslant \int_0^tf(A(\sigma))d\sigma$, then we can find $T(t,s)\geqslant 0$ such that
\begin{equation}\label{EQ2.12}
\int_{t-T(t,s)}^{t}f(A(\sigma ))d\sigma =s-s_{-}
\end{equation}%
in the region $R_{2}$, so it denotes the time it takes for a juvenile to
grow to size $s$ at time $t$ from the minimal size $s_{-}$. Replacing $s$ in $u(t,s)$ with (\ref{EQ2.10}), we can compute formally as follows, assuming that $u$ is a $C^{1}$-function:
\begin{eqnarray*}
& & \frac{d}{dt}u\left(t,C+\int\nolimits_{0}^{t}f(A(\sigma ))d\sigma\right) \\
& = & \partial _{t}u\left(t,C+\int\nolimits_{0}^{t}f(A(\sigma ))d\sigma \right)+f(A(t))\partial
_{s}u\left(t,C+\int\nolimits_{0}^{t}f(A(\sigma ))d\sigma \right) \\
& = & -\mu \left(C+\int\nolimits_{0}^{t}f(A(\sigma ))d\sigma\right) u\left(t,C+\int\nolimits_{0}^{t} f(A(\sigma )) d\sigma \right).
\end{eqnarray*}
This is a separable ODE with respect to $t$. Integration of this equation, and by using the initial distribution and the boundary condition, we obtain the following expression of $u(t,s)$
\begin{equation}\label{EQ2.13}
u(t,s)=\left\{\begin{array}{l}
\displaystyle u_{0}\left( s-\int\nolimits_{0}^{t}f(A(\sigma ))d\sigma \right)
e^{-\int_0^t\mu(s-\int\nolimits_{0}^{t}f(A(\sigma ))d\sigma+\int\nolimits_{0}^{l}f(A(\sigma ))d\sigma ) dl },\\
\qquad\qquad\qquad\qquad\qquad\qquad\qquad\textrm{if } s\geqslant s_{-}+\int_0^t f(A(\sigma))d\sigma,\\
\displaystyle\frac{\beta b(A(t-T(t,s)))}{f(A(t-T(t,s)))}e^{-\int_{t-T(t,s)}^t\mu( s_{-}+\int\nolimits_{t-T(t,s)}^{l} f(A(\sigma )) d\sigma)dl},\\
\qquad\qquad\qquad\qquad\qquad\qquad\qquad\textrm{if } s\leqslant  s_{-}+ \int_0^tf(A(\sigma))d\sigma.
\end{array}\right.
\end{equation}
Whenever $s^{\ast }-s_{-} \leqslant \int_0^tf(A(\sigma))d\sigma$,  we can specifically define $\tau (t):=T(t,s^{\ast })$ as the solution of
\begin{equation}\label{EQ2.14}
\int_{t-\tau (t)}^{t}f(A(\sigma ))d\sigma =s^{\ast }-s_{-}.
\end{equation}
Actually the term $\tau (t)=T(t,s^{\ast })$ represents the time spent by a newborn becoming an adult.

We now assume the mortality function as follows:
\begin{equation*}
\mu (s)=\left\{
\begin{array}{l}
\mu _{A}>0,\text{ if }s\geqslant s^{\ast }, \\
\mu _{J}>0,\text{ if }s\in \lbrack s_{-},s^{\ast }).%
\end{array}%
\right.
\end{equation*}
Then when $s=s^{\ast }$, we have for $t\in[0,t^{\ast}]$,
\begin{equation*}
u(t,s^{\ast })=u_{0}\left(s^{\ast }-\int\nolimits_{0}^{t}f(A(\sigma))d\sigma\right)e^{-\mu _{J}t},
\end{equation*}%
and for $t>t^{\ast}$,
\begin{equation*}
u(t,s^{\ast })=\frac{\beta b(A(t-\tau(t)))}{f(A(t-\tau(t)))}e^{-\mu _{J}\tau (t)}.
\end{equation*}%
Replacing $u(t,s^{\ast })$ back in (\ref{EQ2.8}), we get the model (\ref{EQ2.5}).
 
By differentiating the second equation of (\ref{EQ2.5}) in time, we obtain
\begin{equation*}
\frac{d}{dt}\int\nolimits_{t-\tau (t)}^{t}f(A(\sigma ))d\sigma
=0\Leftrightarrow f(A(t))-f(A(t-\tau (t)))\left( 1-\tau ^{\prime }(t)\right)
=0.
\end{equation*}%
Therefore the state-dependent delay differential equation (\ref{EQ2.6}) is derived.

\begin{remark}
Note that the function $t\rightarrow t-\tau (t)$ is strictly increasing
because
\begin{equation*}
\dfrac{d}{dt}\left( t-\tau (t)\right) =\dfrac{f(A(t))}{f(A(t-\tau (t)))}>0.
\end{equation*}
\end{remark}

We conduct a comparison between the growth function (\ref{EQ2.2}) and the intrinsic function of the growth submodel in the simulator SORTIE. From the model (\ref{EQ2.3}), we only care about the growth of juveniles, so first we assume the radius function of a juvenile
\begin{equation*}
r(t)=\frac{\mathrm{diam_{10}}(t)}{2}
\end{equation*}
where $\mathrm{diam}_{10}$ represents the diameter at 10cm height. We use the following change of variable to define the size $s$ which we are using in the model (\ref{EQ2.1})
\begin{equation}\label{EQ2.15}
s(t):=\ln\frac{r(t)}{r_-},
\end{equation}
where $r_-$ is the minimal radius of the juvenile. We will have
\begin{equation*}
s'(t)=\frac{r'(t)}{r(t)}=f(A(t)).
\end{equation*}
Then the approximation of the derivative of $r(t)$, which describes the growth of the radius, is
\begin{equation}\label{EQ2.16}
\frac{r(t+\Delta t)-r(t)}{\Delta t}=r(t)\cdot\frac{\alpha}{1+\delta A(t)}=r(t)\cdot\frac{\alpha A(t)^{-1}}{\delta+A(t)^{-1}}.
\end{equation}
Take $\Delta t=1$ (one year), then (\ref{EQ2.16}) shows the increase of the radius in one year.

On the other hand, we have the following formula for growth in SORTIE from \citep{Pacala1993, Kobe1995, Pacala1996}:
\begin{equation}\label{EQ2.17}
\mathrm{Annual\ Radius\ Increase}=\mathrm{Radius}\cdot\frac{G_1\cdot \mathrm{GLI}}{\displaystyle\frac{G_1}{G_2}+\mathrm{GLI}},
\end{equation}
where $G_1$ is the asymptotic growth rate at high light and $G_2$ is the slope at 0 or low light. The term GLI (global light index) describes the percentage of light transmitted through tree gaps and perceived by trees, thus is a measure for light. Comparing the two formulas (\ref{EQ2.16}) and (\ref{EQ2.17}), we find that they have the same form, and the parameters $A(t)^{-1},\alpha ,\delta$ correspond to $\mathrm{GLI},G_1,\dfrac{G_1}{G_2}$ respectively. So the choice of the growth function (\ref{EQ2.2}) is reasonable. Plus, this also explains what is size $s$ in our model (\ref{EQ2.1}). By this definition of $s(t)$, we have the minimal size of juveniles $s_-=0$ (as $r(t)=r_-$), and the minimal size of adults $\displaystyle s_*=\ln\frac{r^*}{r_-}$, where $r^*$ is the minimal radius of adults. 

\section{Positive equilibrium for two-species model}\label{AppB}

We compute the positive equilibrium for the system (\ref{EQ3.2}), which is, we compute the solution for the following equations:
\begin{equation}\label{A.1}
\left\{\begin{array}{l}
0=e^{-\mu_{J_1}\tau_1}\beta_1 A_1-\mu_{A_1}A_1,\\
\displaystyle\int_{t-\tau_1}^t\frac{\alpha_1}{1+\delta_1(\zeta_{11}A_1+\zeta_{12}A_2)}d\sigma=s^*-s_-,
\end{array}\right.
\end{equation}
and
\begin{equation}\label{A.2}
\left\{\begin{array}{l}
0=e^{-\mu_{J_2}\tau_2}\beta_2 A_2-\mu_{A_2}A_2,\\
\displaystyle\int_{t-\tau_2}^t\frac{\alpha_2}{1+\delta_2(\zeta_{21}A_1+\zeta_{22}A_2)}d\sigma=s^*-s_-,
\end{array}\right.
\end{equation}
Obviously, $A_1=0$, $A_2=0$ is a trivial equilibrium for the species, in which case we have
\begin{equation*}
\tau_1=\frac{s^*-s_-}{\alpha_1},\ \tau_2=\frac{s^*-s_-}{\alpha_2}.
\end{equation*}
Moreover, we have two "boundary" equilibrium $(\bar{A}_1,0)$ and $(0,\tilde{A}_2)$, where
\begin{equation*}
\bar{A}_1=\frac{1}{\delta_1\zeta_{11}}\left(\frac{\alpha_1}{\mu_{J_1}(s^*-s_-)}\ln\frac{\beta_1}{\mu_{A_1}}-1\right),
\end{equation*}
\begin{equation*}
\bar{\tau}_1=\frac{1}{\mu_{J_1}}\ln\frac{\beta_1}{\mu_{A_1}},\ \bar{\tau}_2=\frac{(s^*-s_-)(1+\delta_2\zeta_{21}\bar{A}_1)}{\alpha_2},
\end{equation*}
and
\begin{equation*}
\tilde{A}_2=\frac{1}{\delta_2\zeta_{22}}\left(\frac{\alpha_2}{\mu_{J_2}(s^*-s_-)}\ln\frac{\beta_2}{\mu_{A_2}}-1\right),
\end{equation*}
\begin{equation*}
\tilde{\tau}_1=\frac{(s^*-s_-)(1+\delta_1\zeta_{12}\tilde{A}_2)}{\alpha_1},\ \tilde{\tau}_2=\frac{1}{\mu_{J_2}}\ln\frac{\beta_2}{\mu_{A_2}}.
\end{equation*}
Now we turn to the positive equilibrium. As $A_1, A_2\neq0$, we solve the first equation in \eqref{A.1} and \eqref{A.2} and get
\begin{equation}\label{A.3}
\tau_1=\frac{1}{\mu_{J_1}}\ln\frac{\beta_1}{\mu_{A_1}},\quad\tau_2=\frac{1}{\mu_{J_2}}\ln\frac{\beta_2}{\mu_{A_2}}
\end{equation}
By the second equation of \eqref{A.1} and \eqref{A.2}, we have
\begin{equation}\label{A.4}
\left\{\begin{array}{l}
\displaystyle\zeta_{11}A_1+\zeta_{12}A_2=\frac{1}{\delta_1}\left(\frac{\alpha_1\tau_1}{s^*-s_-}-1\right),\vspace{0.1cm}\\
\displaystyle\zeta_{21}A_1+\zeta_{22}A_2=\frac{1}{\delta_2}\left(\frac{\alpha_2\tau_2}{s^*-s_-}-1\right)
\end{array}\right.
\end{equation}
We replace $\tau_1$ and $\tau_2$ in \eqref{A.4} by \eqref{A.3}, and we get the following linear equations:
\begin{equation}\label{A.5}
\left\{\begin{array}{l}
\zeta_{11}A_1+\zeta_{12}A_2=\Phi_1,\\
\zeta_{21}A_1+\zeta_{22}A_2=\Phi_2,
\end{array}\right.
\end{equation}
where
\begin{equation*}
\Phi_1:=\frac{1}{\delta_1}\left[\frac{\alpha_1}{\mu_{J_1}(s^*-s_-)}\ln\frac{\beta_1}{\mu_{A_1}}-1\right],\quad\Phi_2:=\frac{1}{\delta_2}\left[\frac{\alpha_2}{\mu_{J_2}(s^*-s_-)}\ln\frac{\beta_2}{\mu_{A_2}}-1\right].
\end{equation*}
First, as we want a positive solution, we need the following conditions:
\begin{equation}\label{A.6}
\Phi_1\geqslant0,\ \Phi_2\geqslant0.
\end{equation}
We solve the equation \eqref{A.5} directly without considering its solvability:
\begin{equation}\label{A.7}
A_1=\frac{\zeta_{22}\Phi_1-\zeta_{12}\Phi_2}{\zeta_{11}\zeta_{22}-\zeta_{12}\zeta_{21}},\quad A_2=\frac{\zeta_{11}\Phi_2-\zeta_{21}\Phi_1}{\zeta_{11}\zeta_{22}-\zeta_{12}\zeta_{21}}.
\end{equation}
In order to have a positive solution, we need the following conditions:
\begin{equation}\label{A.8}
\left\{\begin{array}{l}
\zeta_{11}\zeta_{22}-\zeta_{12}\zeta_{21}>0,\\
\zeta_{22}\Phi_1-\zeta_{12}\Phi_2>0,\\
\zeta_{11}\Phi_2-\zeta_{21}\Phi_1>0,
\end{array}\right.\text{or\ }
\left\{\begin{array}{l}
\zeta_{11}\zeta_{22}-\zeta_{12}\zeta_{21}<0,\\
\zeta_{22}\Phi_1-\zeta_{12}\Phi_2<0,\\
\zeta_{11}\Phi_2-\zeta_{21}\Phi_1<0,
\end{array}\right.
\end{equation}
or in another simplified form
\begin{equation}\label{A.9}
\frac{\zeta_{12}}{\zeta_{22}}<\frac{\Phi_1}{\Phi_2}<\frac{\zeta_{11}}{\zeta_{21}},\text{\quad or\quad}\frac{\zeta_{11}}{\zeta_{21}}<\frac{\Phi_1}{\Phi_2}<\frac{\zeta_{12}}{\zeta_{22}},
\end{equation}

So we have
\begin{lemma}
Under the condition \eqref{A.6} and \eqref{A.9}, the equations \eqref{A.1} and \eqref{A.2} have a positive equilibrium as in \eqref{A.7}.
\end{lemma} 
We check the conditions \eqref{A.6} and \eqref{A.9} for our previous results in Table \ref{TABLE1}-\ref{TABLE3}, and we have
\begin{equation*}
\Phi_1=100.6839>0,\quad\Psi_2=133.4324>0, 
\end{equation*}
\begin{equation*}
\frac{\zeta_{12}}{\zeta_{22}}=0.6,\quad\frac{\Phi_1}{\Phi_2}=0.7546,\quad\frac{\zeta_{11}}{\zeta_{21}}=0.625,
\end{equation*}
which does not satisfy the condition \eqref{A.9}, so there is no positive equilibrium in our previous simulation, and eastern hemlock is disappearing. In order to have a positive equilibrium, we reduce the influence of American beech towards eastern hemlock, namely we lower $\zeta_{21}$ from 1.6 to 1. Then we have
\begin{equation*}
\frac{\zeta_{11}}{\zeta_{21}}=1,
\end{equation*}
which satisfies the condition \eqref{A.9}. And we have the coexistence of both species as is shown in Figure \ref{FIG11} and Figure \ref{FIG15}. 

\end{appendices}

%\bibliographystyle{plain}
%\bibliography{ref} 

\newpage 

\section*{Tables}

\renewcommand{\thetable}{\arabic{table}}
\setcounter{table}{0}

\begin{center}
\begin{table}[H] \centering%
\begin{tabular}{cccc}
\hline
Parameter & Interpretation & value & Reference \\ \hline
$\mu _{J_1}$ & natural mortality rate for juveniles & 0.03 & estimated \\
$\mu _{A_1}$ & natural mortality rate for adults & 0.001 & estimated \\
$\beta_1 $ & birth rate in absence of birth limitation & 2 & estimated \\
$s_{-}$ & minimal size for juvenile & 0 & \citep{Pacala1993} \\
$s^{\ast }$ & minimal size for adult & $\ln 50$ & \citep{Ribbens1994} \\
$\xi_1 $ & parameter in the Ricker type function & 0 & estimated \\
$\tau _{01}$ & time delay of the juveniles present & 121 & estimated \\
& at time 0 to become adults &  &  \\
$\alpha_1 $ & growth rate of juveniles without adults & 0.1709 & computed \\
$\delta_1 $ & parameter describing the descending speed &  &  \\
& of the growth ratewhen adult population increases & 0.1 & estimated \\ 
 \hline
\end{tabular}%
\caption{Parameter values of the best fit for species 1: American beech.}\label{TABLE1}%
%TCIMACRO{\TeXButton{E}{\end{table}}}%
%BeginExpansion
\end{table}%
%EndExpansion
\end{center}

\begin{center}
\begin{table}[H] \centering%
\begin{tabular}{cccc}
\hline
Parameter & Interpretation & value & Reference \\ \hline
$\mu _{J_2}$ & natural mortality rate for juveniles & 0.031 & estimated \\
$\mu _{A_2}$ & natural mortality rate for adults & 0.0037 & estimated \\
$\beta_2 $ & birth rate in absence of birth limitation & 4 & estimated \\
$s_{-}$ & minimal size for juvenile & 0 & \citep{Pacala1993} \\
$s^{\ast }$ & minimal size for adult & $\ln 50$ & \citep{Ribbens1994} \\
$\xi_2 $ & parameter in the Ricker type function & 0 & estimated \\
$\tau _{02}$ & time delay of the juveniles present & 127 & estimated \\
& at time 0 to become adults &  &  \\
$\alpha_2$ & growth rate of juveniles without adults & 0.249 & computed \\
$\delta_2$ & parameter describing the descending speed &  &  \\
& of the growth rate when adult population increases & 0.1 & estimated \\
 \hline
\end{tabular}%
\caption{Parameter values of the best fit for species 2: eastern hemlock.}\label{TABLE2}
\end{table}
\end{center}

\begin{center}
\begin{table}[H] \centering%
\begin{tabular}{cccc}
\hline
Parameter & Interpretation & value & Reference \\ \hline
$\zeta_{11}$ & parameter in the competition term describing &  &  \\
& the intraspecific competition among American beech & 1 & estimated \\
$\zeta_{12}$ & parameter in the competition term describing the &  & \\
& interspecific influence of eastern hemlock on American beech & 0.6 & estimated \\
$\zeta_{21}$ & parameter in the competition term describing the &  & \\
& interspecific influence of American beech on eastern hemlock & 1.6 & estimated \\
$\zeta_{22}$ & parameter in the competition term describing &  & \\
& the intraspecific competition among eastern hemlock & 1 & estimated \\
$\tau_{01}$ & time delay of the American beech juveniles & 201 & estimated \\
& present at time 0 to become adults & &\\
$\tau_{02}$ & time delay of the eastern hemlock juveniles & 208 & estimated \\
& present at time 0 to become adults & &\\
\hline
\end{tabular}
\caption{Other parameter values for two-species model.}
\label{TABLE3}
\end{table}
\end{center}

\end{document}